\documentclass{aa}  

\usepackage{graphicx}
\usepackage{booktabs}
\usepackage{txfonts}
\usepackage[colorlinks,citecolor={blue}]{hyperref}
\newcommand{\jlm}[1]{{\color{cyan}#1}}
\newcommand{\refe}[1]{{\color{black}#1}}

\begin{document}

   \title{The NICER view of Scorpius X-1}


   \author{J. López-Miralles
          \inst{1,2}
          \and
          S. E. Motta \inst{3,4}
          \and
          J.C.A. Miller-Jones \inst{5}
          \and
          J. Homan \inst{6}
          \and
          J. Kajava \inst{7}
          \and
          S. Migliari \inst{1}
          }


   \institute{Aurora Technology for the European Space Agency, ESAC/ESA, Camino Bajo del Castillo s/n, Urb. Villafranca del Castillo, Spain\\ \email{jose.lopezmiralles@ext.esa.int}
        \and      
             Departament d’Astronomía i Astrofísica, Universitat de València, Dr. Moliner 50, 46100, Burjassot (València), Spain
        \and
            INAF–Osservatorio Astronomico di Brera, via E. Bianchi 46, 23807 Merate (LC), Italy
        \and
            Department of Physics, Astrophysics, University of Oxford, Denys Wilkinson Building, Keble Road, OX1 3RH Oxford, UK
        \and
            International Centre for Radio Astronomy Research–Curtin University, GPO Box U1987, Perth, WA 6845, Australia
        \and
            Eureka Scientific, Inc., 2452 Delmer Street, Oakland, CA 94602, USA
        \and
            Serco for the European Space Agency, ESAC/ESA, Camino Bajo del Castillo s/n, Urb. Villafranca del Castillo, Spain
             }

   \date{Received April 25, 2025; accepted July 28, 2025}

 
  \abstract
   {The Neutron Star X-ray binary Scorpius X-1 is one of the brightest Z-type sources in our Galaxy, showing frequent periods of flaring activity and different types of relativistic outflows. Observations with RXTE have shown that the strongest X-ray variability appears in the transition from/to the flaring state. During this transition, it has been proposed that the appearance of two particular types of quasi-periodic oscillations (QPOs) might be connected with the ejection of the so-called ultra-relativistic flows.}
   {In this paper, we present an analysis of the first NICER observations of Scorpius X-1 obtained during a multi-wavelenght campaign conducted in February 2019, in order to characterise the properties of QPOs in this source as the system evolves through its various accretion states.} 
   {We compute a light-curve and a Hardness-Intensity diagram to track the evolution of the source spectral properties, while we investigate the X-ray time variability with a Dynamical Power Density Spectrum. To trace the temporal evolution of QPOs, we segment the dataset into shorter, continuous intervals, and compute and fit the averaged power density spectrum for each interval.}
   {Our analysis shows that the overall behaviour of the source is consistent with the literature; strong QPOs around 6 Hz are detected on the normal branch, while transitions to/from the flaring branch --occurring over timescales of a few hundreds of seconds-- are characterised by rapid, weaker quasi-periodic variability reaching frequencies up to 15 Hz. Despite limited statistical significance, we also identify faint, transient timing features above 20 Hz, which occasionally coexist with the prominent 6 Hz QPOs. Although tentative, the existence of these timing features in the NICER data is crucial for interpreting the simultaneous radio observations from the same multi-wavelength campaign, potentially reinforcing the connection between the ejection of relativistic outflows and the accretion states in Scorpius X-1.}
   {}


   \maketitle
%

\section{Introduction}

Neutron Star Low-mass X-ray binaries (NS LMXBs) are stellar binary systems in which the compact object (CO; the accretor) is a NS orbiting a low-mass companion (the donor). Based on their spectral and timing properties, they are classified in two main categories: Z-type sources and atoll-type sources \citep{hasinger89}, depending on the characteristic shape that NS LMXBs trace in the X-ray colour-colour diagram (CD) or in the Hardness-Intensity diagram (HID). In particular, Z-type sources are the most luminous, accreting near or above the Eddington rate. In the aforementioned diagrams, they draw a characteristic Z-like shape where we can distinguish three separate branches: the horizontal branch (HB), the normal branch (NB) and the flaring branch (FB). A hard apex (HA) separates the HB and the NB,
and a soft apex (SA) separates the NB and the FB. These branches represent specific accretion states of the system, which may be crossed on short (hours or less) time-scales \citep{darias14}. Moreover, all known Z-type NS LMXBs are persistent radio emitting sources. 

The radio luminosity of Z-source varies as a function of the state in the CD/HID \citep{penninx88,hjellming90a,hjellming90b}, which is stronger on the HB (or high NB), and quenched in the FB (or low NB). This variability is physically connected to specific changes in the accretion flow, similar to the disc-jet coupling observed in LMXBs hosting a stellar-mass black hole \citep[BH; e.g. ][]{migliari06}. 

The power density spectrum (PDS) of Z-type NSs also changes from one accretion state to another \citep[i.e.,][]{mendez99}, showing different type of quasi-periodic oscillations (QPOs) that can be classified in two main categories: kHz QPOs, detected isolated or in pairs above $500$~Hz \citep[i.e.,][]{klis89}, and low-frequency QPOs, typically appearing below $50$~Hz. The latter are further divided in three different types, depending on the branch of the CD/HID where they manifest \citep[e.g. ][]{klis89}: horizontal branch oscillations (HBOs), normal branch oscillations (NBOs) and flaring branch oscillations (FBOs). These are considered to be the NS equivalent of the so-called type C, type B and type A QPOs found in BH systems \citep[][]{ingram19}.

The X-ray binary (XRB) Scorpius X-1 (hereafter, Sco X-1) was discovered in 1962 \citep{giacconi62}, and it is the brightest extrasolar X-ray source in the sky, located at a distance of $2.33$~kpc \citep{dr3}. It is classified as a LMXB with an orbital period of 18.9 h \citep{got75}, containing a later-than-K4 spectral type companion star (0.28M$_{\odot}$<M<0.70M$_{\odot}$) and a weakly magnetised NS with mass <1.73 M$_{\odot}$ \citep{mata15}. 
Sco X-1 is one of the handful of permanent Z-type sources in our Galaxy (together with GX 349+2, GX 340+0, GX 17+2, GX 5-1, Cyg X-2, GX 13+1, and the peculiar system Cir X-1), and it is known to show extended radio luminous jets \citep[i.e., radio lobes;][]{fomalont01a}.

Over the past few decades, several X-ray observations of Sco X-1 have investigated the timing properties of the source \citep[see e.g.,][and references therein]{titarchuk14}. Using EXOSAT, \cite{mid86} first discovered a $6$~Hz QPO, which was later shown to vary from 10 to 20 Hz on short time scales during an active state \citep{pried86,vanderklis87,dieters2000}. \cite{klis96} reported the discovery of 45 Hz QPOs in the RXTE data, besides (sub)millisecond oscillations (the so-called kHz QPOs), correlated with the frequency of the 6-20 Hz QPOs in the NB/FB. Some years later, \cite{casella06} showed the first resolved rapid transition from a FBO to an NBO, with a monotonic (smooth) increase of the QPO centroid frequency from 4.5-7 Hz (in the NB) to 6–25 Hz (in the FB).

In the radio band, \cite{fomalont01a} showed that the radio structure of Sco X-1 is composed by a radio core and two symmetric radio lobes. In \cite{fomalont01b}, the authors suggested for the first time the existence of relativistic flows travelling along the jets --later dubbed ultra relativistic flows (URFs)--, inferred by correlating X-ray flares in the core with subsequent radio flares in the approaching and receding lobes. This type of fast flows, whose origin and physical properties are still unclear, have been also observed in at least two other XRBs, namely Cir X-1 (\citealt{fender04,tudose08}, but see also \citealt{miller12}) and SS 433  (\citealt{migliari05}, but see also \citealt{miller09}). More recently, \cite{motta19} re-analysed the RXTE data of Sco X-1 and suggested that this type of outflows are related with the appearance of a particular class of QPO, while radio emitting outflows were associated with flat-top broad-band noise components. In particular, the authors found evidence that URF ejections occur when both an NBO and an HBO are present in the X-ray power spectrum. This might indicate, at least tentatively, that the ejection of URFs is connected with specific changes in the accretion flow.

From 21 to 25 February 2019, an extensive multi-wavelength campaign was organised to monitor Sco X-1 with different instruments, providing the largest simultaneous coverage at all possible wavelengths with radio/VLBI (VLBA+EVN+LBA, rotating through each array as the Earth rotated, to provide continuous coverage), radio/VLA, optical/IR (SALT, VLT, NOT and TNG), X-ray (NICER, Chandra, XMM-Newton) and $\gamma$-ray (INTEGRAL) observations. In the X-ray energy range, the large collecting area of the NICER  observatory \citep{gendreau12}, together with its ability to handle very high count rates, provided the highest quality data of the X-ray instruments involved in the campaign.


In this paper, we present an analysis of the timing properties of Sco X-1 using NICER observations performed during the February 2019 multi-wavelength campaign. We will mainly focus on the different type of LF QPOs that appear in the dataset, and on how the characteristics of these QPOs (specially the central frequency) change along the position in the HID.  This constitutes the first comprehensive analysis of Sco X-1 based on NICER X-ray observations.

The paper is organised as follows: in Sec.~\ref{obsdata}, we describe the NICER observations and the methodology we follow to produce a light-curve, the HID and the (dynamical) PDS. In Sec.~\ref{results}, we present the main results of the timing analysis. Finally, in Sec.~\ref{discussion}, we discuss our results and we draw our main conclusions.

\section{Observations and data analysis}
\label{obsdata}

\subsection{MAXI}

We first produced a light-curve and a HID using MAXI data taken from 01-02-2019 to 31-03-2019 (from 58515 to 58573 MJD), using the MAXI/GSC on-demand web interface\footnote{http://maxi.riken.jp/mxondem/} \citep{maxi09}. We analysed data extracted within a circular region of $1.6^{\circ{}}$ radius around the source position (244.979455, -15.640283), while the background region was extracted in a concentric ring with outer radius of $3^{\circ{}}$. The bin size is 1 hour (i.e., 0.04 days). Light-curves were extracted in two energy bands: 2-10 keV (A) and 10-20 keV (B), and the hardness-ratio (HR) was defined as HR=B/A.

\subsection{NICER}

The NICER dataset is composed by three different observations identified with proposal number 110803010 (Obs-1; from 58536.450 to 58536.978 MJD), 1108030111 (Obs-2; from 58537.026 to 58537.940 MJD) and 1108030112 (Obs-3; from 58537.991 to 58538.777 MJD). For each observation, we used the unfiltered merged Measurement/Power Unit (MPU) Science data (available under the path /xti/event\_cl/) and performed the standard cleaning using the high-level screening task \textsc{nicerclean} (Heasoft v6.28). Data analysis (including light-curve and PDS extraction) was performed using the General High-energy Aperiodic Timing Software (\textsc{ghats})\footnote{See  http://www.brera.inaf.it/utenti/belloni/GHATS\_Package/Home.html.}. This software, which is an interactive Timing package working on IDL/GDL, has been extensively validated for X-ray timing with RXTE and other missions, including NICER observations \citep[see e.g.,][]{belloni05,motta17,motta19,bhargava19,lm23,belloni24}.

\subsubsection{Light-curve and Hardness-Intensity diagram}
\label{sec:lchr}

We produced a light-curve and a HID with a time bin resolution of 13s. Time gaps were removed from the data, as well as instrumental drops at the beginning of each segment. We extracted  the count rate in the energy bands A$=[0.5-2]$~keV, B$=[2-15]$~keV, and we defined the HR as HR=B/A. The HID was produced by plotting the HR vs. the intensity (i.e., the accumulated count rate in the two energy bands, A+B).

\subsubsection{(Dynamical) Power Density Spectra}
\label{sec:dynamical}

We rebinned the NICER high-resolution Science mode data in time ($\times~10^4$) to obtain a Nyquist frequency of 1250 Hz. Using Fast Fourier Transform (FFT) techniques \citep{klis89}, we produced PDS for continuous 13s-long intervals and we averaged them to obtain a PDS for the whole dataset (including the three observations). We show this PDS in the traditional $\nu, P_{\nu} $ representation, where we applied the Leahy normalisation \citep{leahy83}. PDS were fitted with the \textsc{xspec} package using a one-to-one energy-frequency conversion and a unity response matrix. This allowed us to fit a power versus frequency spectrum as if it was a flux versus energy spectrum. Along this work, we always consider a multi-Lorentzian model plus a power law component to take into account the contribution of the Poisson noise \citep[see e.g.,][]{belloni02}. To show how the PDS changes with time, we also computed a dynamical PDS following similar techniques ($P_{\nu}$ in the time vs. frequency plane). 

To search for fainter and/or short-lived timing features, we divided the whole dataset in continuous shorter segments, taking advantage of the location of the light-curve gaps. For each of these segments, we produced a collection of PDS by averaging 30 consecutive 13s-long bins (i.e., $390$~s), covering the total time length of the three observations. Before averaging, PDS sets containing less than 15 bins were automatically discarded. In an attempt to increase the strength of the variability, for this part of the analysis we also removed photons in the energy range below 750~eV.

\subsection{XMM-Newton: the brightest observation ever}

XMM-Newton observed Sco X-1 between 21 February 2021 and 22 February 2021 for a total exposure time of approximately 100 ks (rev. 3517, ObsId 0831790901). The EPIC-pn camera was operated in Burst mode with the THIN filter, resulting in the highest count rate ever measured with the XMM-Newton observatory \refe{(above $7\times10^4$ counts/s)}. However, this extraordinarily high rate, which was above the operational limits of the EPIC-pn instrument, led to multiple resets due to the extremely high work load of the on-board electronics (FIFO overflows), besides the quasi-periodic triggering of the so-called counting mode due to telemetry saturation (with an activation period of approximately 18s). As a result, the timing information in the science files was corrupted, both in terms of integration time for spectral products and of GTI correction of rates in the resulting light-curves. Therefore, we conclude that the XMM-Newton data taken during February 2019 is not appropriate for timing studies of Sco X-1, and we will not consider it for this paper.

\section{Results}
\label{results}

\begin{figure}
	\includegraphics[width=0.97\linewidth]{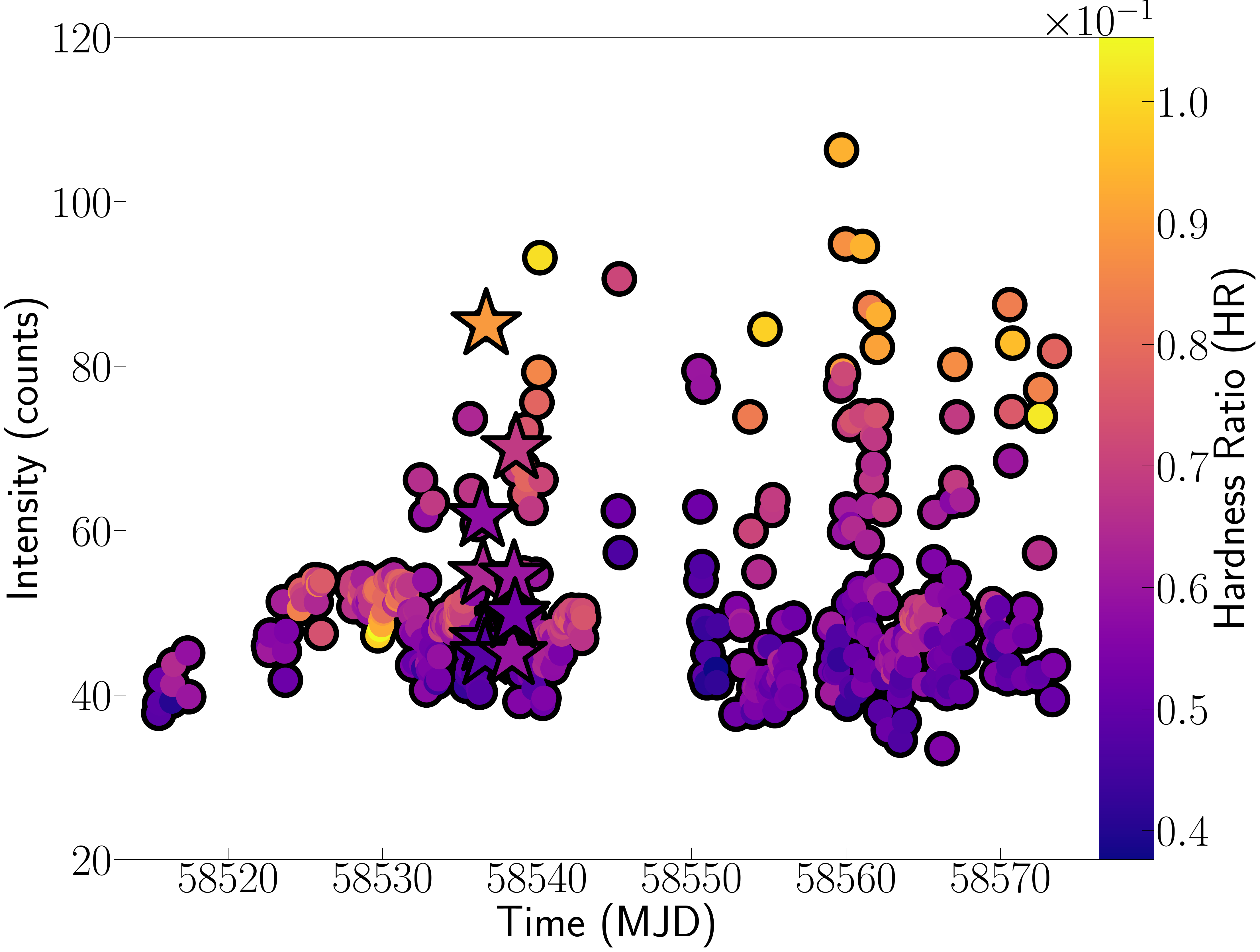}
	\includegraphics[width=\linewidth]{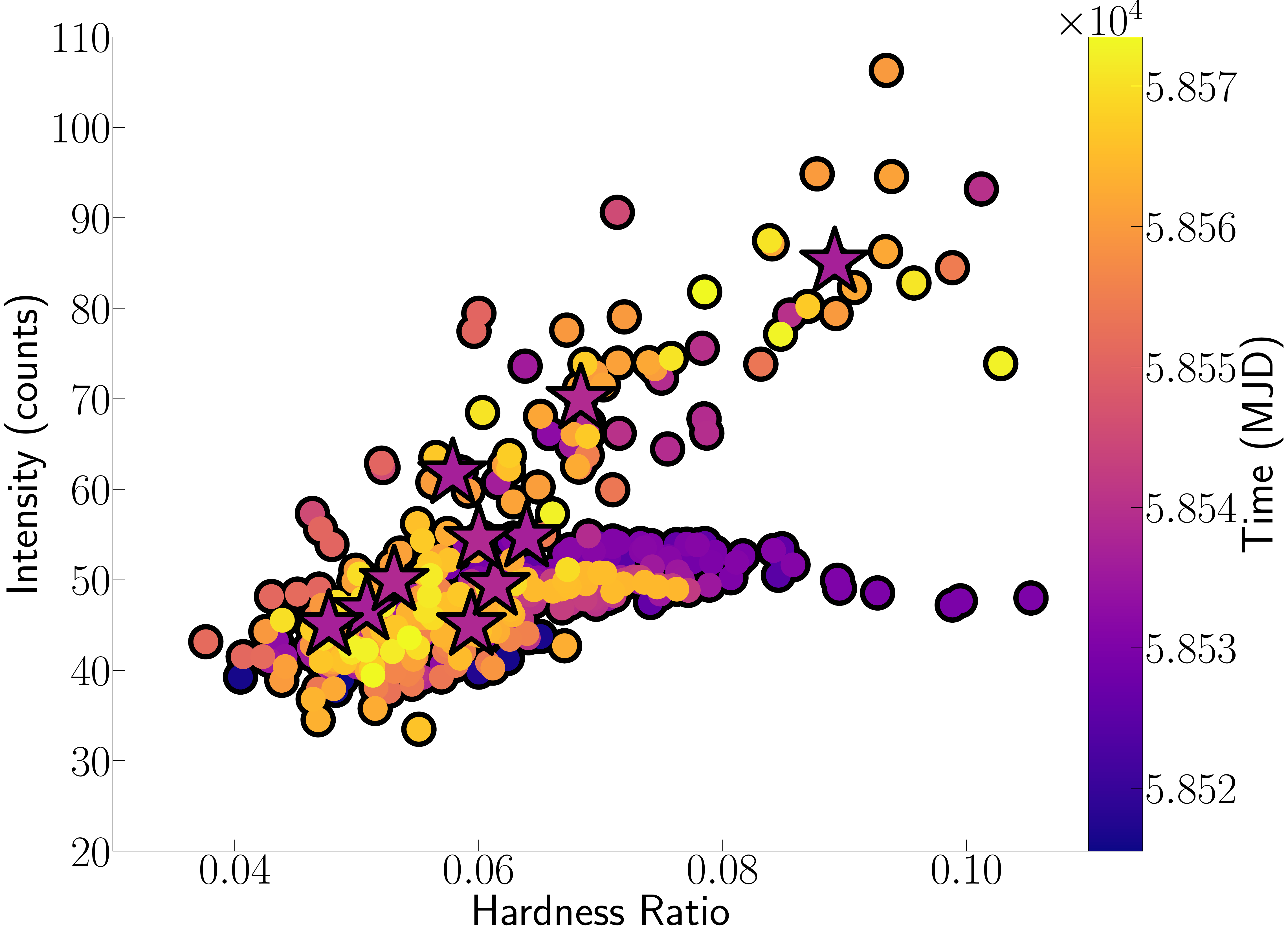}
\caption{Light-curve (top) and HID (bottom) of Sco X-1 using MAXI data taken from 01-02-2019 to 31-03-2019 with a time resolution of 0.04 days. The light-curve is colour-coded with the HR=B/A, and the HID with the time (in MJD). Data taken from 22-02-2019 to 24-02-2019, corresponding to the multi-wavelength campaign, is represented with star markers.}
    \label{maxilc}
\end{figure}

Figure~\ref{maxilc} shows the MAXI light-curve coloured by the HR (A+B, in units of counts; top) and the HID coloured by time in MJD (B/A vs. A+B; bottom). The light-curve shown in Fig.~\ref{maxilc} (top) reveals that the source experiences frequent flaring activity along the two months of MAXI continuous monitoring, where the HID (bottom) shows the characteristic pattern of a NS Z-source, with a FB, a NB and a HB (from top to bottom). In both panels, data collected simultaneously with NICER during the February 2019 multi-wavelength monitoring (from 58536.4 to 58538.7~MJD) is represented with star markers. This small subset of points ($\sim 1$~hr time bins) shows two distinct important incursions into the FB, the first one at $58536.7$~MJD and the second one at $ 58538.6$~MJD. However, the information provided by the light-curve in this time range is limited, since no MAXI observations were collected from $58536.7$ to $58538.4$~MJD.

\begin{figure*}
	\includegraphics[width=\linewidth]{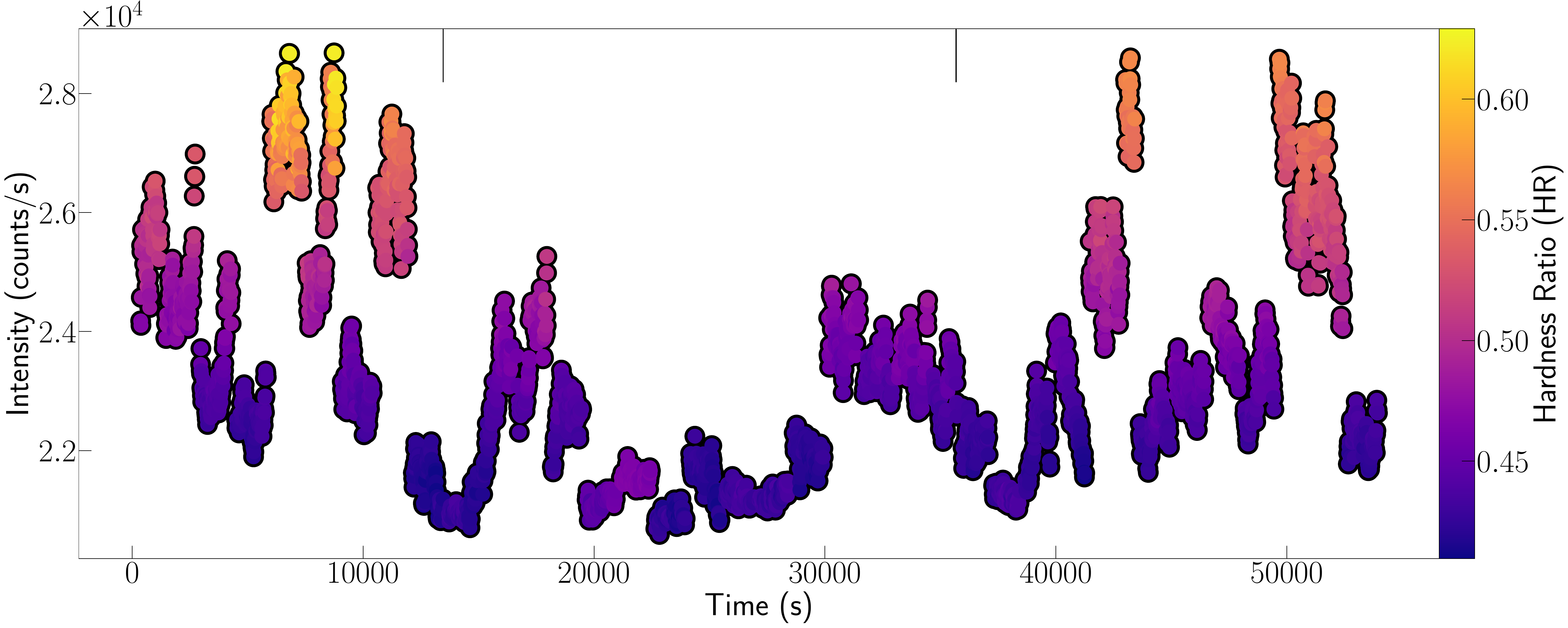}
\caption{Light-curve (in units of counts/s) for the full dataset after removing time gaps and instrumental drops. The two division marks at the top of the plot separates the three NICER observations. Color scale represents the HR as defined in Sec.~\ref{sec:lchr}.}
    \label{lc}
\end{figure*}

\begin{figure*}
	\includegraphics[width=\linewidth]{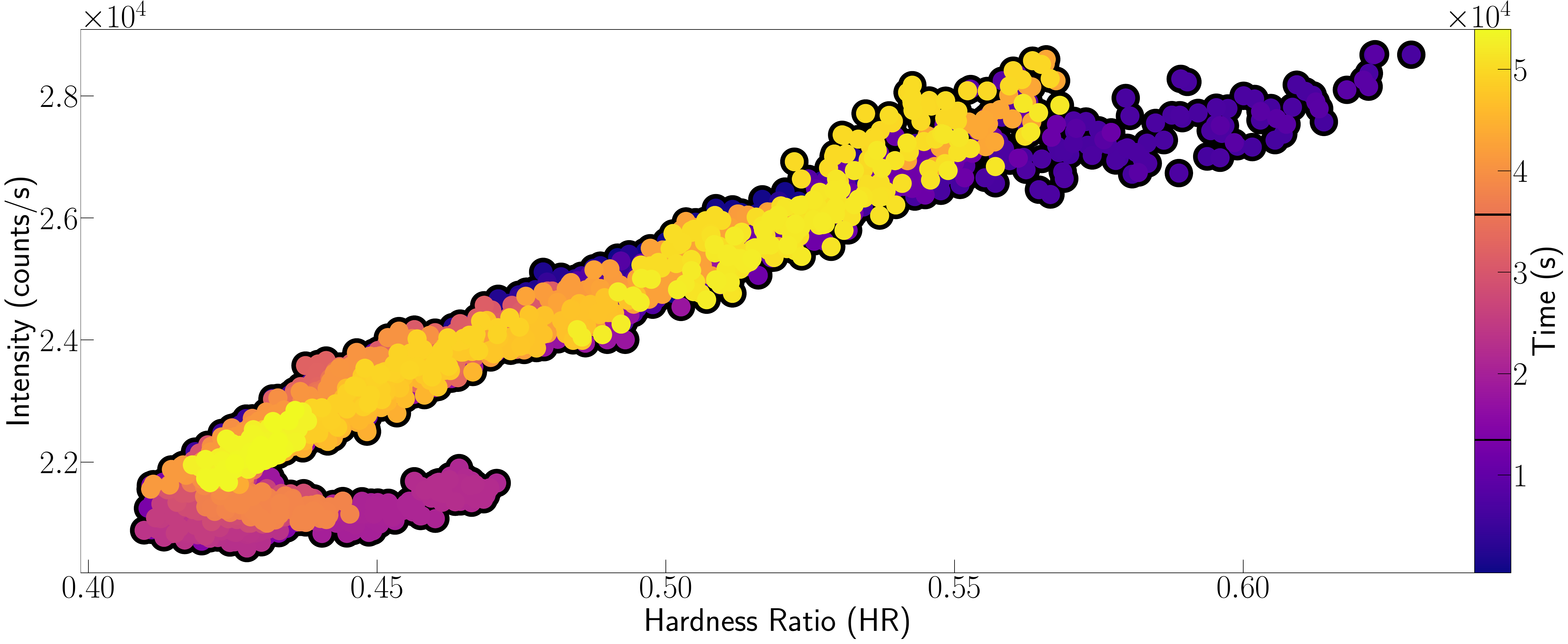}

\caption{Hardness-Intensity diagram for the entire dataset. Color scale represents the relative time in seconds after removing time gaps and instrumental drops (as shown in the x-axis of Fig.~\ref{lc}). The two division marks in the colorbar show the limit of each NICER observation.}
    \label{ccd}
\end{figure*}

\begin{figure}
	\includegraphics[width=\linewidth]{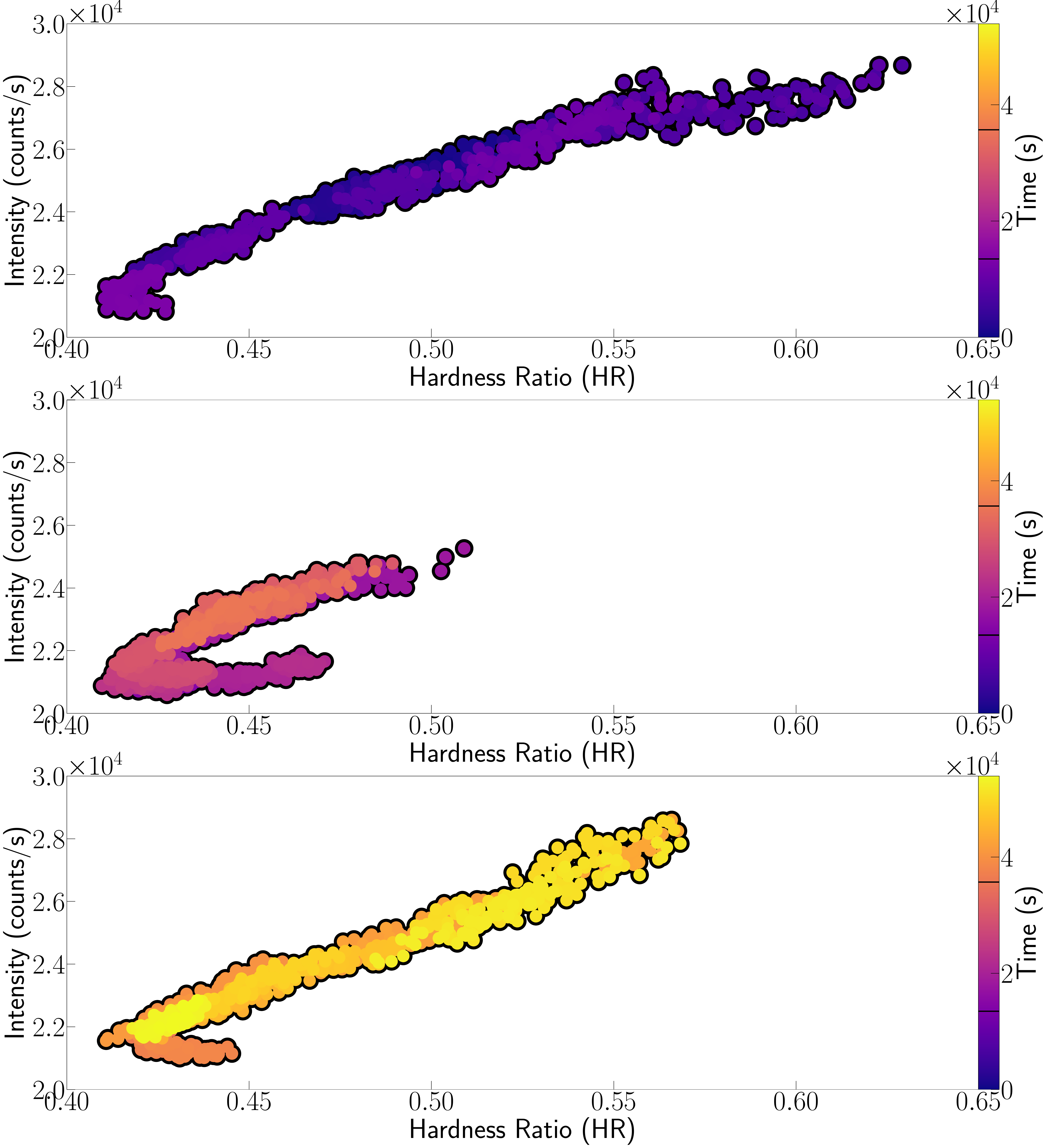}

\caption{Same as Fig.~\ref{ccd} for Obs-1 (top), Obs-2 (middle) and Obs-3 (bottom). In each panel, the time in the colorbar corresponds to the total relative time of the full dataset as displayed in the x-axis of Fig.~\ref{lc}.}
    \label{ccd-multi}
\end{figure}

The NICER light-curve of Fig.~\ref{lc} shows these two big flares with much more time resolution, where the x-axis corresponds to the relative time after removing all time gaps and instrumental drops, and the colormap represents the HR as defined in Sec.~\ref{sec:lchr}. Figs.~\ref{ccd}-\ref{ccd-multi} show respectively the HID for the full dataset, and for each individual observation. 

During the flaring state, the source is highly variable, showing back and forth rapid transitions. It is also interesting to note that, despite that the main flares show approximately the same count rate ($\sim 28000$~counts/s), the HR is higher during the first incursion into the FB. In the HID of Fig.~\ref{ccd}, this creates a fishhook-like pattern in the top part of the branch, with a harder --and longer-- branch corresponding to the first burst, and a softer branch corresponding to the second one. In the NICER data, the NB is generally not well resolved, and the transition to the HB can only be slightly deduced in Obs-2 (see Fig.~\ref{ccd-multi}, central plot).

\begin{figure}

	\includegraphics[width=\linewidth]{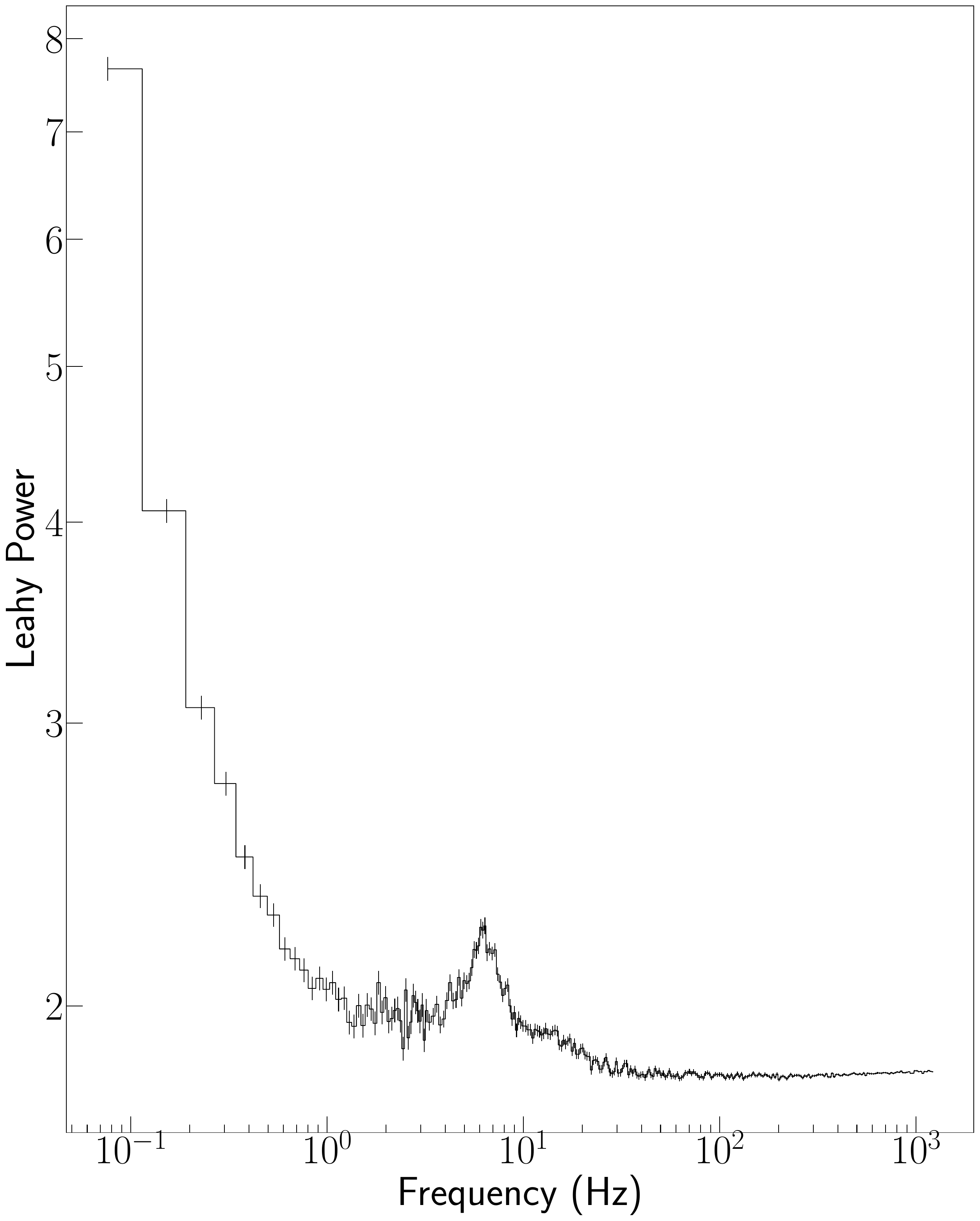}
    \caption{Power density spectrum for the entire NICER dataset, including Obs-1, Obs-2 and Obs-3. Broad-band red-noise appears below $\sim30$ Hz, while a strong QPO is observed at $6.35\pm 0.06$~Hz. At higher frequencies, power is dominated by Poisson noise.}
    \label{pds-total}
\end{figure}

\begin{figure*}

	\includegraphics[width=\linewidth]{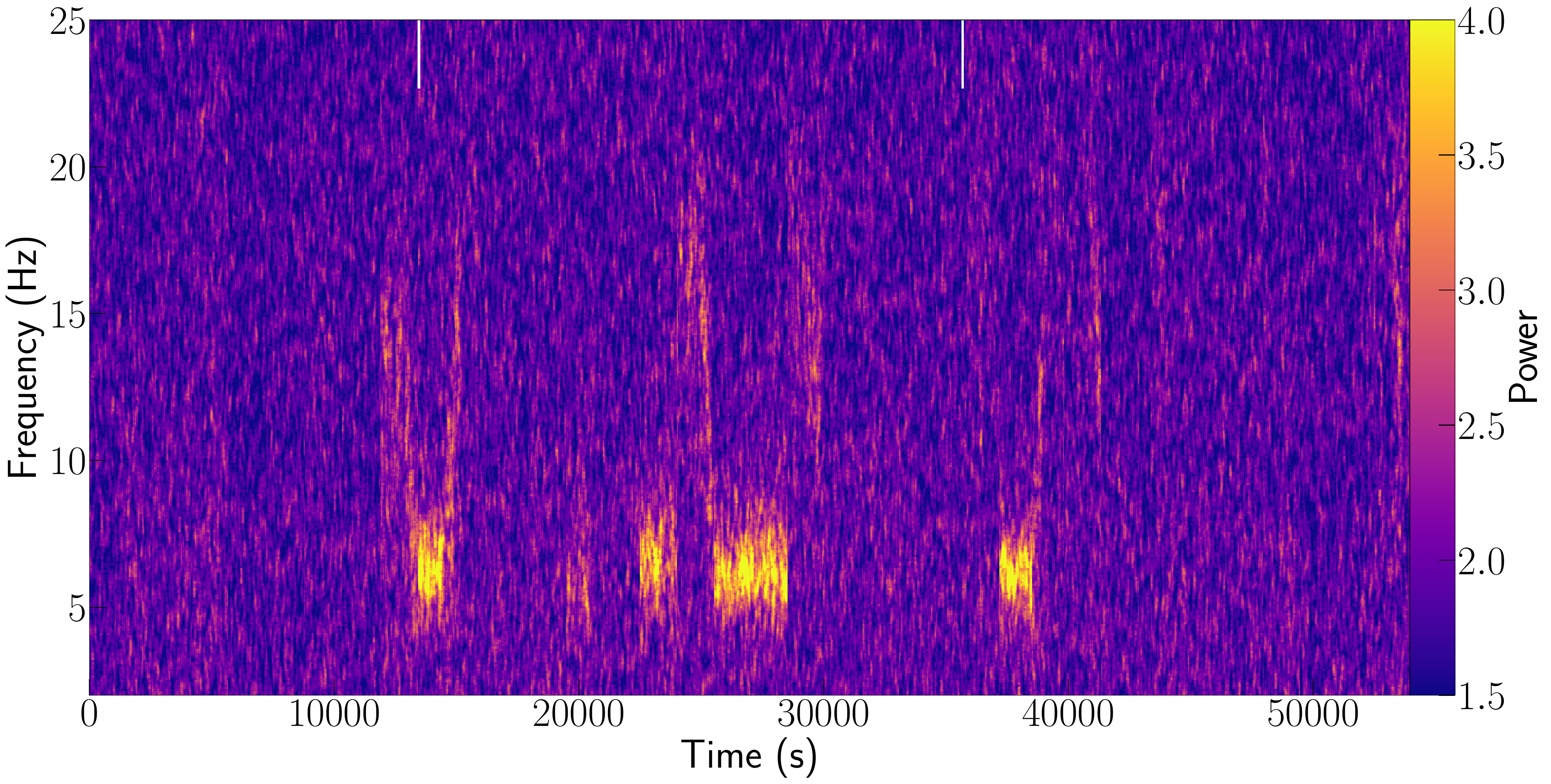}
    \caption{Dynamical power density spectrum for the entire NICER dataset, including Obs-1, Obs-2 and Obs-3. The plot shows strong signal below 10 Hz and weaker fast variability between 10 Hz and 20 Hz. The two division marks at the top of the plot separates the three NICER observations.}
    \label{dynamical}
\end{figure*}

The averaged PDS shown in Fig.~\ref{pds-total} is dominated by a $\sim 6$~Hz QPO, revealing low-frequency red noise below 1 Hz and flat-top noise above this frequency (aside from the QPO). As described in Sec.~\ref{sec:dynamical}, we fitted the PDS with a five-Lorentzian model plus a power-law component to account for the contribution of the Poisson noise. Using this model, the central frequency of the QPO yields $6.35\pm 0.06$~Hz ($\chi^2_{\nu}$=176/156). This feature is the byproduct of the strong variability shown in the dynamical PDS of Fig.~\ref{dynamical} (in yellow-white), which also shows weaker variability in the 6-20 Hz frequency range. The strongest variability signal appears when the source is in the quiet state (near the light-curve minima at $2\times 10^4$~counts/s), while rapid --and fainter-- variability up to 20~Hz precedes/follows the source flaring activity ($> 2.2\times 10^4$~counts/s).

In order to characterise these transitions along the HID, in the rest of this section we will report the results of the analysis described in the second paragraph of Sec.~\ref{sec:dynamical}. For this purpose, we have computed averaged PDS every $\sim400$~s, which were fitted with a multi-Lorentzian+power-law statistical model in \textsc{xspec}. As described in App.~\ref{ref:app-tables}, each PDS is defined with a unique token that identifies the PDS number, the light-curve segment (i.e. the continuous chunks of data between light-curve gaps) and the position of the PDS in the segment, in this order.

In this paper, we assume the following convection to claim a QPO discovery: we consider narrow features (Q>2, where Q is the QPO quality factor\footnote{We define the quality factor $Q=\nu_0/2\Delta$, where $\nu_0$ is the central frequency of the QPO, and $\Delta$ is the half width at half maximum.}) with sufficient statistical significance ($3\sigma$), although a few exceptions of this rule are also considered. On the one hand, when the Q-factor is Q$\le 2$ and the statistical significance is $> 3\sigma$, QPOs are referred as \textit{broad} or  \textit{blurred} features. On the other hand, when the statistical significance is $\le 3\sigma$ but Q>2, QPOs are referred as \textit{faint} or \textit{hint} narrow features. This nomenclature is relevant to follow the dynamical evolution of the statistically significant QPOs in this source, while providing evidence on the existence of other features which are however difficult to resolve in our dataset. The results of the analysis for each individual PDS are reported in the tables of App.~\ref{ref:app-tables}.

\subsection{Observation 1 (110803010)}

The main characteristics of the QPOs found in Obs-1 (central frequency, Q-factor and statistical significance in number of sigmas) are displayed on Tab.~\ref{tab01}. 

In this first observation, the source spends most of the time in the FB, showing several transitions from the low FB to the high FB. As shown in the dynamical PDS of Fig.~\ref{dynamical}, no relevant aperiodic variability is expected during this state, and PDS are mostly dominated by broad-band noise. The strongest signal appears on segment \#08 (see Tab.~\ref{app:tab1}), when the source moves from the FB towards the NB. This transition is characterised by the presence of a well-resolved FBO, which decreases in frequency as the source approaches the SA. In some of the PDS of the last segment (\#290800, \#320803), we observe a second component at lower frequencies, possibly due to changes of the FBO frequency in short time scales (this second component was, however, difficult to constrain in the model due to low signal-to-noise at these low frequencies).

Besides the LF QPOs near the SA, in two PDS distributed along the FB (i.e., \#010001, \#070202) we also found very narrow QPO-like features in the frequency range between 30~Hz and 40~Hz (see Tab.~\ref{app:tab1}). Although these features are statistically significant according to the adopoted convention (i.e., $>3\sigma$), these narrow QPOs were difficult to constrain in the models so we do not report them as part of the main findings of Tab.~\ref{tab01}.

\begin{table*}
	\footnotesize
	\centering
	\caption{Centroid frequency (Hz), Q-factor and statistical significance (in number of $\sigma$) of the QPOs found in Obs-1. The last column shows the goodness of fit for the multi-lorentzian model employed to fit each PDS.}
	\label{tab01}
	\begin{tabular}{lccccccccccc} 
		\toprule
            \multicolumn{1}{c}{} & \multicolumn{3}{c}{\textbf{NBO}} & \multicolumn{3}{c}{\textbf{FBO}} & \multicolumn{3}{c}{\textbf{HBO}} & \multicolumn{1}{c}{} \\
            \cmidrule(rl){2-4} \cmidrule(rl){5-7} \cmidrule(rl){8-10} 
            \textbf{Id} & {Freq. (Hz)} & {Q factor} & {$\sigma$}  & {Freq. (Hz)} & {Q factor} & {$\sigma$}&  {Freq. (Hz)} & {Q factor} & {$\sigma$} & \bf $\chi_{\nu}^2$(dof) \\

		\midrule
            

            \bf220601 & - & - & - & $12.34^{+0.68}_{-0.65}$  & 3.46 & 3.28  & - & - & - &  1.01(122) \\

		
            \bf290800 & $8.49^{+1.14}_{-0.92}$ & 0.95 & 4.03 & $14.58^{+0.25}_{-0.25}$ & 6.85 & 3.04 &- &- &-& 0.80(128)  \\
		
            \bf300801 & - & - & -  &$11.83^{+0.48}_{-0.50}$  & 1.60 & 9.64 & - & - &  - & 0.85(122) \\

           \bf310802 & - & - & -  &$9.75^{+0.50}_{-0.49}$  & 1.12 & 10.90  & - & - &  - & 0.99(122) \\
            
           \bf320803 & $4.50^{+0.19}_{-0.16}$ & 7.14 & 3.03 &$7.41^{+0.25}_{-0.24}$  & 3.15 & 5.38 & - & - & - &  1.04(119)  \\

		\bottomrule

	\end{tabular}

\end{table*}

\subsection{Observation 2 (110803011)}

The main characteristics of the QPOs found in Obs-2 are shown in Tab.~\ref{tab11}. This NICER observation shows high variability, where approximately half of the PDS reveal some kind of QPO feature. 

The three PDS of segment \#00 are dominated by a strong NBO (with statistical significance $>12\sigma$), that shifts slightly in frequency from 6.34 Hz (PDS \#000000) to 7.67 Hz (PDS \#020002) as the source moves from the NB towards the SA (see also Tab.~\ref{app:tab2}).

In this observation, we can distinguish two main incursions into the FB; the first in segment \#01 ($t\sim18000$~s in the light-curve of Fig.~\ref{lc}) and the second one in segment \#10 ($t\sim32000$~s in the light-curve of Fig.~\ref{lc}). There is also a fast transition into the low-FB in segment \#07 , where the dynamical PDS shows rapid back and forth variability in the frequency range 6-20 Hz (see Fig.~\ref{dynamical}). The first incursion into the FB is characterised by the presence of an FBO in the first PDS of segment \#01 (with central frequency 13.6 Hz), which is the consequence of the shift of the NBO of segment \#00 towards higher frequencies. In the FB, PDS show broad-band noise components with no aperiodic features. On segment \#04, the flaring activity ceases and the source returns to the NB, where NBOs reappear. As the HR increases and the source moves towards the HA, the NBO smears in the PDS noise. On segment \#05, the source is in the HA, where we could resolve a single isolated HBO at $33.65$~Hz (Q$\sim$9 and 3.7$\sigma$), which is the only example of this type of QPO that we could find in the whole NICER dataset. Every other attempt to fit an HBO-like feature resulted in very low statistical significance (see e.g., PDS \#00000 and PDS \#180600 in Tab.~\ref{tab11}). However, given the importance of this kind of QPOs for the purpose of this paper, we report our results even when the statistics is not formally relevant.

When the source moves again towards the SA, the strong NBO reappears. In segment \#07, the source shows a rapid incursion into the low FB that triggers an FBO in the frequency range 14-17 Hz, which turns back into a strong NBO when the source returns to the NB in segment \#08 and segment \#09. From segment \#10 onwards, the source moves into the FB and the FBOs are detected again in the PDS, despite the source being in the mid-low FB state (see e.g. PDS \#361200 in Tab.~\ref{app:tab2}).

\begin{table*}
	\footnotesize
	\centering
	\caption{Centroid frequency (Hz), Q-factor and statistical significance (in number of $\sigma$) of the QPOs found in Obs-2. The last column shows the goodness of fit for the multi-lorentzian model employed to fit each PDS. Numbers in blue represent our attempt to fit HBO-like features together with the strong NBOs at 6 Hz, which led to low statistical significance.}
	\label{tab11}
	\begin{tabular}{lccccccccccc} 
		\toprule
            \multicolumn{1}{c}{} & \multicolumn{3}{c}{\textbf{NBO}} & \multicolumn{3}{c}{\textbf{FBO}} & \multicolumn{3}{c}{\textbf{HBO}} & \multicolumn{1}{c}{} \\
            \cmidrule(rl){2-4} \cmidrule(rl){5-7} \cmidrule(rl){8-10} 
            \textbf{Id} & {Freq. (Hz)} & {Q factor} & {$\sigma$}  & {Freq. (Hz)} & {Q factor} & {$\sigma$}&  {Freq. (Hz)} & {Q factor} & {$\sigma$} & \bf $\chi_{\nu}^2$(dof) \\
       
              \midrule
              \bf000000 & $6.34^{+0.08}_{-0.08}$ & 3.39 & 15.27 & - & - & - & \jlm{$21.57^{+3.11}_{-1.93}$} & \jlm{4.14} & \jlm{1.60} &  1.02(119)\\

              \bf010001 & $6.40^{+0.08}_{-0.08}$ & 3.79 & 13.41 & - & - & - & - & - & - &  0.69(122)\\

              \bf020002 & $7.67^{+0.31}_{-0.30}$ & 1.27 & 12.81 & - & - & - & - & - & - &  1.00(122)\\
               
              \bf030100 & - & - & - & $13.58^{+0.70}_{-0.78}$ & 1.20 & 8.03 & - & - & - & 1.08(122)\\

              \bf 060200 & $4.88^{+0.50}_{-0.46}$ & 1.37 & 4.25 & - & - & - & - & - & - &  1.29(122)\\
                 
              \bf120400 & $5.84^{+0.22}_{-0.21}$ & 2.57 & 6.19 & - & - & - & - & - & - &  1.05(122)\\

              \bf130401 &  $5.81^{+0.38}_{-0.34}$ & 1.20 & 7.88 & - & - & - & - & - & - &  0.87(122)\\

              \bf150500 & - & - & - & - & - & - & $33.65^{+0.72}_{-0.77}$ & 8.95 & 3.67 &  0.85(122)\\

              \bf180600 & $6.80^{+0.14}_{-0.14}$ & 2.65 & 11.95 & - & - & - & \jlm{$19.50^{+0.82}_{-1.20}$} & \jlm{9.24} & \jlm{1.84} & 0.96(119)  \\

              \bf190601 & $6.68^{+0.10}_{-0.10}$ & 3.20 & 12.77 & - & - & - & - & - & - &  0.86(122)\\

              \bf200602 & $7.32^{+0.26}_{-0.27}$ & 1.57 & 10.96 & - & - & - & - & - & - &  0.92(122)\\

              \bf210700 & - & - & - & $16.59^{+0.45}_{-0.44}$ & 4.26 & 3.70 & - & - & - & 0.97(119)\\

              \bf220701 & - & - & - & $17.05^{+0.33}_{-0.33}$ & 3.87 & 6.62 & - & - & - &  1.02(119)\\

              \bf230702 & - & - & - & $12.86^{+0.45}_{-0.48}$ & 1.54 & 12.46 & - & - & - &  1.14(120)\\

              \bf240800 & $6.27^{+0.12}_{-0.12}$ & 2.81 & 11.55 & - & - & - & - & - & - &  1.11(122)\\

              \bf250801 & $6.13^{+0.14}_{-0.14}$ & 2.52 & 11.07 & - & - & - & - & - & - &  1.16(122)\\

              \bf260802 & $6.25^{+0.07}_{-0.07}$ & 4.08 & 14.00 & - & - & - & - & - & - & 1.26(122)\\

              \bf270900 & $6.46^{+0.11}_{-0.11}$ & 3.02 & 12.76 & - & - & - & - & - & - &  1.35(122)\\
               
              \bf280901 & $6.43^{+0.09}_{-0.09}$ & 2.98 & 13.77 & - & - & - & - & - & - & 1.25(122)\\
               
              \bf290902 & $6.20^{+0.13}_{-0.12}$ & 2.56 & 13.02 & - & - & - & - & - & - &  0.90(122)\\

              \bf301000 & - & - & - & $16.17^{+0.65}_{-0.70}$ & 2.06 & 7.82 & - & - & - &  1.19(122)\\
 
              \bf311001 & - & - & - & $14.22^{+0.65}_{-0.60}$ & 1.98 & 7.03 & - & - & - &  0.93(122)\\
               
              \bf321002 & - & - & - & $13.62^{+0.31}_{-0.32}$ & 2.36 & 10.19 & - & - & - &  0.89(122)\\
                              

              \bf361200 & - & - & - & $15.80^{+1.68}_{-1.69}$ & 2.05 & 4.02 & - & - & - &  0.97(121)\\

		\bottomrule

	\end{tabular}

\end{table*}

\subsection{Observation 3 (110803012)}

The main characteristics of the QPOs found in Obs-3 are shown in Tab.~\ref{tab21}. In this observation, the source is predominantly active (as it is the case of Obs-1), showing two main incursions into the NB: the first one in segment \#01 ($t\sim43000$~s in the light-curve of Fig.~\ref{lc}), and the second one at the end of the observation, in segment \#12 ($t\sim52000$~s in the light-curve of Fig.~\ref{lc}). 

During the first incursion, 3 PDS show strong NBOs at $\sim6$~Hz (i.e., PDS \#030100, PDS \#040101 and \#PDS 050102). In PDS \#040101, we fitted a second component above 20 Hz, for which we got a borderline detection with low statistical significance ($2\sigma$). This corresponds to the higher significance that we could get in the three observations when attempting to fit an HBO-like component together with the strong NBOs at 6 Hz.

In PDS \#060200, a blurred FBO (Q=0.9) at $\sim9$~Hz marks the transition from the NB to the FB. During the flaring state, PDS are dominated by broad-band noise, but we also detected 3 FBO features in PDS \#110302, PDS \#160600, and PDS \#230801. The last three PDS of the dataset (i.e., \#341200, \#351201, \#361202) represent the second transition to the NB and the end of the flaring state. These PDS are characterised by well-resolved FBOs above 14 Hz.

\begin{table*}
	\footnotesize
	\centering
	\caption{Centroid frequency (Hz), Q-factor and statistical significance (in number of $\sigma$) of the QPOs found in Obs-3. The last column shows the goodness of fit for the multi-lorentzian model employed to fit the PDS. Numbers in blue represent our attempt to fit an HBO-like feature together with the strong NBOs at 6 Hz, which led to low statistical significance. }
	\label{tab21}
	\begin{tabular}{lccccccccccc} 
		\toprule
            \multicolumn{1}{c}{} & \multicolumn{3}{c}{\textbf{NBO}} & \multicolumn{3}{c}{\textbf{FBO}} & \multicolumn{3}{c}{\textbf{HBO}} & \multicolumn{1}{c}{} \\
            \cmidrule(rl){2-4} \cmidrule(rl){5-7} \cmidrule(rl){8-10} 
            \textbf{Id} & {Freq. (Hz)} & {Q factor} & {$\sigma$}  & {Freq. (Hz)} & {Q factor} & {$\sigma$}&  {Freq. (Hz)} & {Q factor} & {$\sigma$} & \bf $\chi_{\nu}^2$(dof) \\
       
              \midrule
                        
              \bf030100 & $6.30^{+0.08}_{-0.09}$ & 3.66 & 13.0 & - & - & - & - & - & - &  0.95(122)\\

              \bf040101 & $6.18^{+0.09}_{-0.09}$ & 3.64 & 11.98 & - & - & - & \jlm{$23.01^{+0.75}_{-0.72}$} & \jlm{$11.51^{\dagger}$} & \jlm{2.00} &  0.95(120)\\

              \bf050102 & $6.17^{+0.10}_{-0.10}$ & 3.35 & 12.23 & - & - & - & - & - & - &  1.13(122)\\

               \bf060200 & -  & - & - & $9.24^{+0.64}_{-0.68}$ & 0.90 & 10.02 & - & - & - &  1.05(122)\\


              \bf110302 & - & - & - & $16.28^{+0.70}_{-0.67}$ & 1.69 & 7.66 & - & - & - &  1.04(122)\\

              \bf160600 & - & - & - & $17.76^{+0.63}_{-1.66}$ & 3.18 & 3.07 & - & - & - &  1.00(122)\\

              \bf 230801 & - & - & - & $6.83^{+0.32}_{-0.35}$ & 3.33 & 3.62 & - & - & - &  0.90(122)\\


              \bf341200 & - & - & - & $17.94^{+0.94}_{-1.21}$ & 2.10 & 3.96 & - & - & - &  0.98(122)\\

              \bf351201 & - & - & - & $17.79^{+1.05}_{-0.84}$ & 2.45 & 4.02 & - & - & - &  0.83(122)\\

              \bf361202 & - & - & - & $14.62^{+0.75}_{-0.69}$ & 1.39 & 8.11 & - & - & - &  0.86(122)\\
		\bottomrule

	\end{tabular}

\begin{scriptsize}
\begin{flushleft}
\dag The width of the QPO is not well constrained so it was frozen before fitting the PDS.
\end{flushleft}
\end{scriptsize}		
	
\end{table*}

\section{Discussion}
\label{discussion}

In this paper, we present the analysis of the X-ray fast time variability properties of the NS LMXB Sco X-1 as observed by NICER.  This monitoring is part of a large multi-wavelength campaign performed on Sco X-1 over almost four full days in February 2019, when NICER provided the highest quality data among all the X-ray instruments. 

We considered three NICER observations for a total exposure of $\sim55$~ks, each of which was splitted in continuous shorter segments, taking advantage of the light-curve gaps. In each of these segments, we computed averaged PDS by stacking 13~s long bins every $\sim 400$~s. 

Approximately half of the 115 PDS we inspected show some kind of QPO feature, while the rest are dominated by broad-band noise components. We observed the following: 20 PDS containing NBOs, 20 PDS showing FBOs, and 1 PDS showing a statistically significant HBO (plus 1 faint detection in Obs-3). We also observed at least another distinct type of QPO that is more difficult to classify according to the above categories, which appeared during the flaring state with no obvious preferential location in the FB. These features are the 3 narrow QPOs detected in the frequency range 30-40~Hz in Obs-1 and Obs-2 (see Tab.~\ref{app:tab1}-\ref{app:tab2}), and above 60 Hz in Obs-3 (see Tab.~\ref{app:tab3}). Although these features are marginally significant and difficult to constrain in the models (so they might correspond to spurious variability), we will briefly discuss them in this section.

\begin{figure*}
	\includegraphics[width=\linewidth]{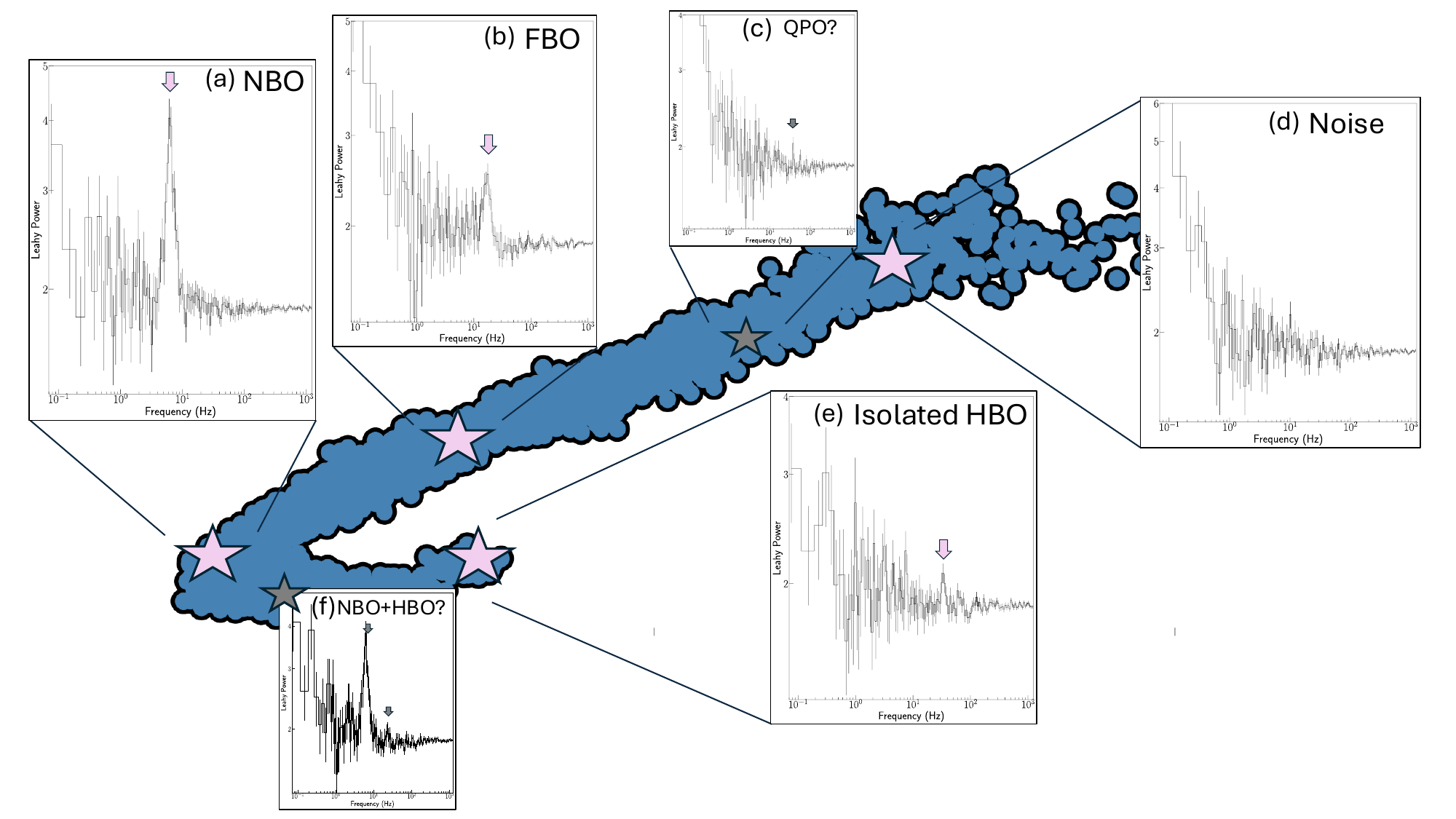} 
    \caption{Fiducial PDS for different spectral states in the HID of Fig.~\ref{ccd}. Bigger panels with pink stars show the main timing features observed in the NICER dataset: the NBOs in the NB (PDS \#010001 of Obs-2), the FBOs in the transition from the NB to the FB (PDS \#220701 of Obs-2), the isolated HBO near the HA (PDS \#150500 of Obs-2) and the broad-band red noise observed in the FB (PDS \#130401 of Obs-3). The two smaller panels with gray stars show less common and tentative features, as the narrow QPOs discovered in the mid FB (PDS \#010001 of Obs-1), or the pair NBO+HBO in the NB (PDS \#040101 of Obs-3).}
    \label{qpo-hid}
\end{figure*}

In order to draw a coherent picture of the source behaviour, Fig.~\ref{qpo-hid} summarises the quasi-periodic variability content of the data in the framework of the HID. In the figure, we show a strong 6 Hz NBO (sub-panel (a), PDS \#010001 of Obs-2, see Tab.~\ref{app:tab2}), an FBO during the transition to the FB (sub-panel (b), PDS \#220701 of Obs-2, see Tab.~\ref{app:tab2}), a narrow QPO-like feature appearing during the flaring state (sub-panel (c), PDS \#010001 of Obs-1, see Tab.~\ref{app:tab1}), broad-band red noise (sub-panel (d), PDS \#130401 of Obs-3, see Tab.~\ref{app:tab3}), the only isolated HBO that we found in our analysis (sub-panel (e), PDS \#150500 of Obs-2, see Tab.~\ref{app:tab2}), and the tentative pair NBO+HBO found in the NB, where the HBO is not statistically significant and thus it is only classified as a hint detection (sub-panel (f), PDS \#040101 of Obs-3, see Tab.~\ref{app:tab3}).

\subsection{Normal-to-flaring branch transitions}
\begin{figure}

	\includegraphics[width=\linewidth]{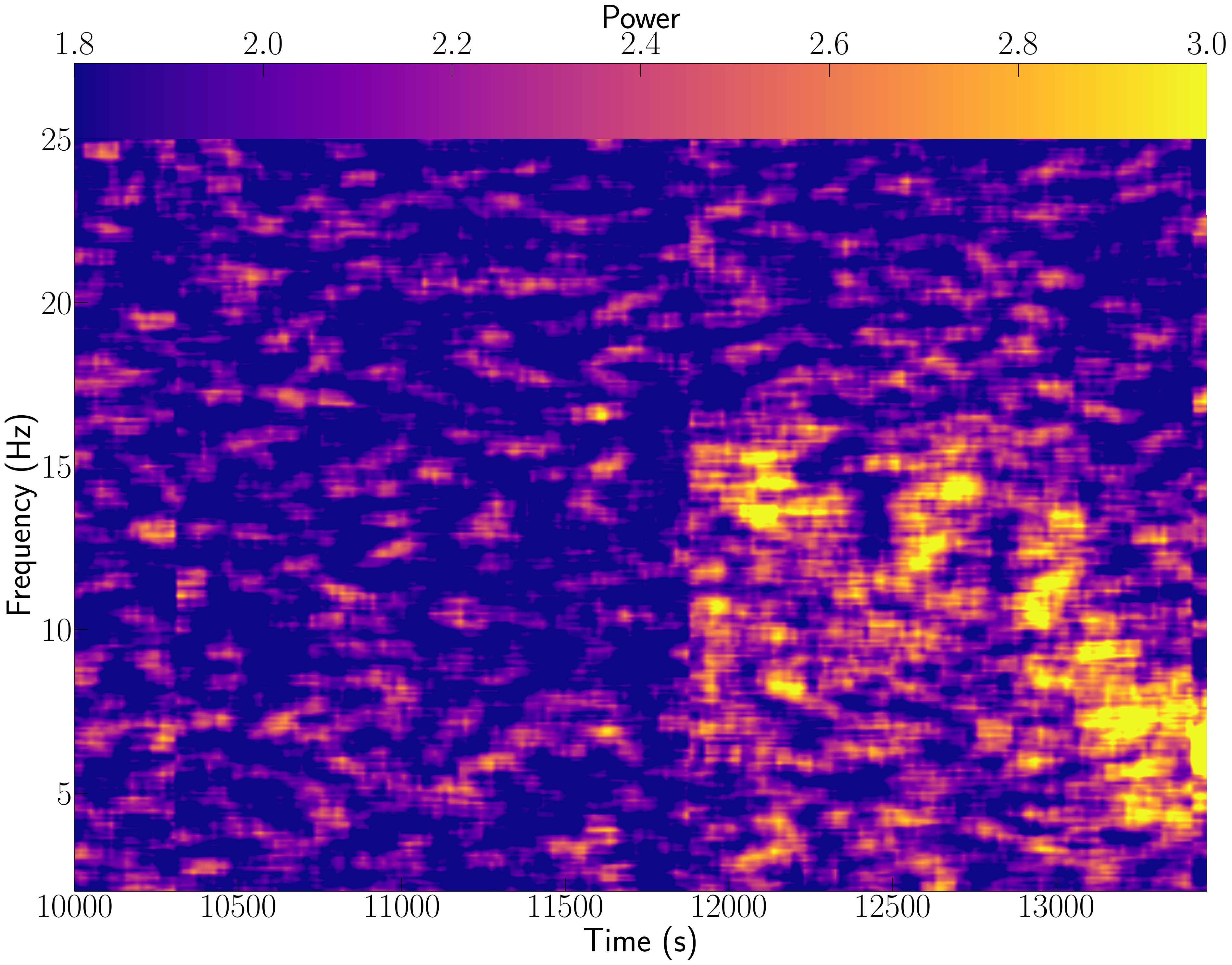}
    \caption{Dynamical power density spectrum for segments \#07 and \#08 of Obs-1. A smooth FB to NB transition is resolved in segment \#08. This corresponds to the unique transition which is fully resolved within one single segment, as all other examples in Obs-2 and Obs-3 coincide with a data gap. Signal smoothly decreases from 15 Hz to 6 Hz in a few hundreds of seconds.}
    \label{dynamical2}
\end{figure}

The overall behavior of the source is consistent with what has been reported in the literature. PDS in the NB are characterised by strong NBOs at $\sim 6$~Hz  \citep[first discovered by][]{mid86}, while all display broad-band red noise as the source moves along the FB. As shown in the dynamical PDS of Fig.~\ref{dynamical}, the transition from the NB to the FB is distinguished by the shift of the NBO towards higher frequencies \citep[or by the shift of the FBO towards lower frequencies in the inverse transition, see e.g. ][]{pried86,vanderklis87,dieters2000,casella06}, while NBOs disappear as the source moves towards the HA (see e.g. PDS \#140402 in Tab.~\ref{app:tab2}). As shown in Fig.~\ref{dynamical2}, the time scale at which this transition takes place is of the order of a few hundred of seconds, in agreement with the results reported in the literature for RXTE \citep{casella06}.

We have found tentative evidence that these transitions might also occur at a much higher rate. This is suggested, for example, by the presence of a statistically significant QPO observed in PDS \#230801 of Obs-3 (see Tab.~\ref{app:tab3}). We classified this QPO as an FBO because it is located in the mid FB, but its central frequency is fully compatible with the strong NBOs of the source, which is lower than the characteristic frequency of the FBOs appearing during the transition from/to the flaring state. Thus, we speculate that the presence of this QPO in the FB may indicate a very rapid excursion of the source into the NB, faster than the typical time scales of the spectral transitions in Sco X-1.

\subsection{Narrow QPOs in the Flaring Branch}

The nature of the narrow QPOs detected in the FB above 30 Hz is unclear. 
The central frequency of these features varies from 30-40 Hz in Obs-1 and Obs-2, to >60 Hz in Obs-3 (see one example in Fig.~\ref{qpo-hid} for PDS \#010001 of Obs-1). Similar features were observed in RXTE observations of Sco X-1 up to $\sim$70 Hz (Motta et al. in prep.). While appearing in the FB, these features show certain similarities with the HBOs seen at lower frequencies. Something similar occurs in BH systems, where type-C QPOs (the HBO equivalent in BHs) are observed in the soft state \citep[see e.g.][]{motta12,franchini17}. The features we observed in the FB might thus be related with the HBOs seen in the HB. In particular, as it has been noted in the literature \citep[see][]{Stella1998, motta17}, such features could be the second harmonic of an undetected fundamental frequency. However, the small number of events, together with the marginal significance of these detections and the difficulty to constrain the properties of such narrow, low-amplitude features, limit the robustness of any conclusion that we could draw on their nature. Therefore, based on the available data, we cannot exclude the possibility that these features are spurious detections.

\subsection{Horizontal Branch Oscillations and URFs}

Given its tentative connection with the ejection of URFs, the existence of HBOs --specially together with an NBO-- should be carefully addressed. The fact that NBO+HBO pairs are less common in the NICER dataset than in RXTE \citep[][]{motta19} is to be expected. Due to the soft response of the NICER detector (0.1-10 keV), HBOs are harder to detect compared to, for instance, the PCA instrument onboard RXTE, which  reached up to 40 keV. Nevertheless, given the relevance of HBOs in the study of relativistic ejections in Sco X-1, along with the existence of an isolated HBO resolved at $34$~Hz, we decided to report in this work any aperiodic feature that shows in this frequency range, albeit not formally statistically significant. It is important to note that, aside from NICER, no other X-ray instrument provides a time resolution high enough to assess this frequency range, so this information --although tentative-- may be relevant for comparison with the radio observations of Sco X-1.

\begin{figure}
	\includegraphics[width=\linewidth]{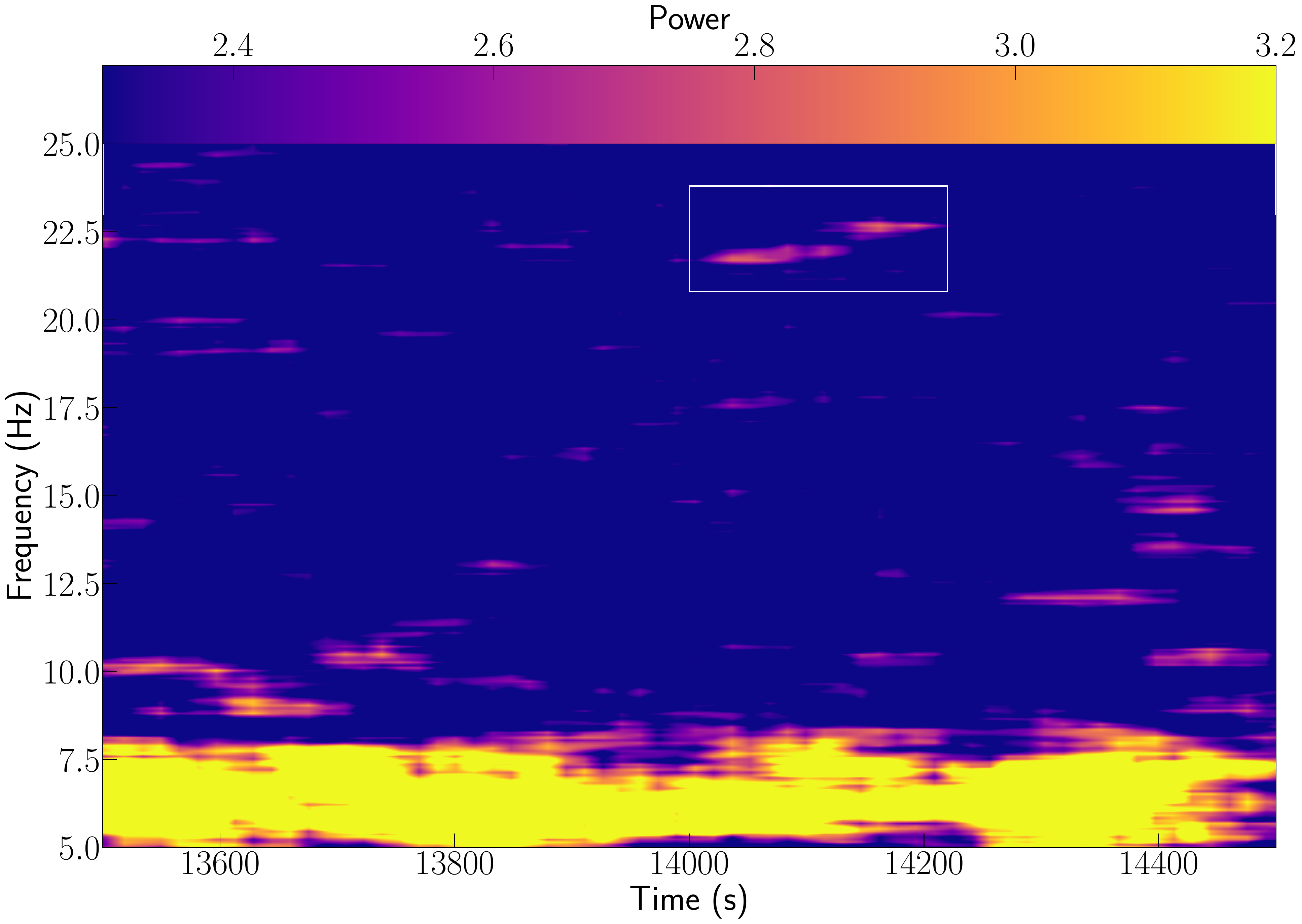}
    \caption{Dynamical power density spectrum for segment \#00 of Obs-2. Transient fast variability is observed above 20 Hz (white box), together with the strong signal around 6 Hz. This variability is compatible with the features reported in Tabs.~\ref{tab11}-\ref{tab21} for Obs-2 and Obs-3, respectively.}
    \label{dynamical3}
\end{figure}

These faint HBO-like features seem to be present together with NBOs in Obs-2 and Obs-3 above 20 Hz, with a maximum significance of $2\sigma$ in PDS \#040101 of Obs-3 (see blue numbers in Tab.~\ref{tab11}-\ref{tab21}). Since we expect HBOs to be transient features, we have closely inspected the dynamical PDS of Fig.~\ref{dynamical} in the segments where our analysis evidences such faint features at higher frequencies (together with an NBO). As shown in Fig.~\ref{dynamical3}, we observe some faint variability in segment \#00 of Obs-2 above 20 Hz, which is compatible with the hint features reported in Tab.~\ref{tab11}-\ref{tab21} for the averaged PDS. We thus consider that the existence of this transient fast variability in the dynamical PDS at the expected frequencies provides more strength to the results discussed in this section.

In \cite{klis97} and \cite{motta19}, all HBOs show approximately the same central frequency ($\sim45$~Hz), while in NICER these features seem to be more scattered at lower frequencies, ranging from 20 Hz (the faint detections coupled to NBOs) to 35 Hz (the single isolated HBO in Obs-3). Due to this fact, we question whether these borderline QPOs (especially the ones at lower frequencies, close to $20$~Hz) could represent the sub-harmonic of an undetected fundamental oscillation, which may be submerged in the Poissonian noise at higher frequencies 
\citep[see e.g.][]{motta17}. However, since HBOs are transient and may be short-lived (see Fig.~\ref{dynamical3}), it is also possible that the characteristic frequency of the QPO varies over time in a wider frequency range. If this is the case, then the NICER characteristics (low or no sensitivity to high energy photons) may be the reason why we do not observe these features at the expected higher frequencies. 

\cite{motta19} showed that the launch of relativistic jets in Sco X-1 is tightly connected with the appearance of certain specific timing features in the PDS of the source, and in particular of an NBO (or the type-B QPO in BH XRBs), generally considered the jet QPO (among others, see \citealt{migliari05}). \refe{Clearly, the fact that NBOs and type-B QPOS may be associated to jets is mostly circumstantial evidence, as the connection between such QPOs and the jet is still unclear and largely debated (as are the possible jet launching mechanisms).}

\cite{motta19} found that the ejection of the (invisible) URFs seems to be associated not only with the appearance of an NBO in the PDS, but also of an HBO at the highest end of the frequency range spanned by such a feature. This implies that the accretion flow is thought to reach as close as possible to the NS surface. Hence, the simultaneous appearance of the NBO + HBO pair indicates that not only an outflow was launched --an URF, as deduced from the radio data-- but that it is launched when the accretion flow is closest to the central NS. Such a configuration has been speculated long over, but to date no better smoking gun has been found than the appearance of a NBO + HBO pair. 

Although the statistics is poor, our results show that the tentative NBO+HBO PDS we found in our data appear at the expected location in the HID, i.e. when the URF is expected to be launched based on \cite{motta19}. Interestingly, preliminary analysis of the radio data taken as part of the same observing campaign shows a number of URF ejections (Motta el al., in prep.), all launched at times consistent with the appearance of the NBO+HBO. Therefore, despite the poor statistical significance of the NICER HBO+NBO detections, we argue that our findings provide further support --at least qualitatively--  to the connection between the accretion event responsible for the appearence of NBO+HBO pairs, and the launch of URFs in Sco X-1.

\begin{acknowledgements}
The project that gave rise to these results received the support of a fellowship from ”La Caixa” Foundation (ID 100010434). The fellowship code is \texttt{LCF/BQ/DR19/11740030}. J.L.M acknowledges additional support from \texttt{MICIU/AEI/10.13039/501100011033} and from Ministerio de Ciencia e Innovación (MCIN) within the Plan de Recuperación, Transformación y Resiliencia del Gobierno de España through the project \texttt{ASFAE/2022/005}.
\end{acknowledgements}

%
   \bibliographystyle{aa} 
   \bibliography{biblio} 
%
\begin{appendix}

\section{Tables}
\label{ref:app-tables}

The results of the timing analysis described in Sec.~\ref{sec:dynamical} are shown in Tab.~\ref{app:tab1} for Obs-1, Tab.~\ref{app:tab2} for Obs-2 and Tab.~\ref{app:tab3} for Obs-3. In these tables, each PDS is characterised with a unique token that identifies the PDS number, the light-curve segment (i.e. the continuous chunks of data between light-curve gaps) and the position of the PDS in the segment, in this order. The tables also show the specific timing events resolved in the PDS, the central frequency of the LF QPOs and the relative location of these bins in the HID. Since in these observations the FB is the longest branch in the diagram, in order to follow the evolution of the source in this state we arbitrarily sub-divided the FB in three different frames: low FB, mid FB and high FB.  Although the subset of bins that defines each PDS is not scattered in the HID, this division might be ambiguous. For these specific cases, we also define two intermediate states: low-mid FB and mid-high FB. 

\begin{table*}
	\footnotesize
	\centering
	\caption{Timing features found in the PDS extracted for Obs-1. In the column QPO, we refer to any feature which cannot be classified in the three main categories (i.e., NBO, FBO, HBO). These features are generally faint (i.e. barely significant) and difficult to constrain in the model; we report them here for completeness, but they are not shown as part of the main tables of the paper.}
	\label{app:tab1}
	\begin{tabular}{lcccccc} 
        \hline
	\bf Id  & \bf Timing event & \bf NBO (Hz) & \bf FBO (Hz) & \bf HBO (Hz) & \bf QPO (Hz) & \bf HID \\
       
		\hline
            \textbf{000000} &  Broad-band noise & - & - & - &- & FB (mid) \\
            \textbf{010001} &  Narrow QPO (?) & - & - & - & $38.20$ & FB (mid) \\
            \hline
            \bf 020100 &  Broad-band noise & - & - & - & - & FB (mid) \\
            \bf 030101 &  Broad-band noise & - & - & - & - & FB (mid) \\
            \bf 040102 &  Broad-band noise & - & - & - & - &FB (mid) \\
            \hline
            \bf 050200 &  Broad-band noise & - & - & - & - &FB (low) \\
            \bf 060201 &  Broad-band noise & - & - & - & - &FB (low) \\
            \bf070202 &  Narrow QPO (?)   & - &  - & - & 34.86 &FB (low) \\
            \bf080203 &  Broad-band noise & - & - & - & - &FB (mid) \\
            \hline
            \bf090300 &  Broad-band noise & - & - & - &- &FB (low) \\
            \bf100301 &  Broad-band noise & - & - & - & - &FB (low) \\
            \bf110302 &  Broad-band noise & - & - & - &- &FB (low) \\
            \bf120303 &  Broad-band noise & - & - & - &- &FB (low) \\
            \hline
            \bf130400 &  Broad-band noise & - & - & - &- &FB (high) \\
            \bf140401 &  Broad-band noise & - & - & - &- &FB (high) \\
            \bf150402 &  Broad-band noise & - & - & - &- &FB (high) \\
            \bf160403 &  Broad-band noise & - & - & - &- &FB (high) \\
            \hline
            \bf170500 &  Broad-band noise & - & - & - &- &FB (mid) \\
            \bf180501 &  Broad-band noise & - & - & - &- &FB (mid) \\
            \bf190502 &  Broad-band noise & - & - & - &- &FB (mid-high) \\
            \bf200503 &  Broad-band noise & - & - & - &- &FB (high) \\            
           \hline

            \bf 210600 &  Broad-band noise & - & - & - & - &FB (low) \\
            \bf220601 &  FBO & - & $12.34$ & - & - &FB (low) \\
            \bf230602 &  Broad-band noise & - & - & - & - &FB (low) \\
            \bf240603 &  Broad-band noise & - & - & - & - &FB (low) \\                  
            \hline
            \bf250700 &  Broad-band noise & - & - & - & -&FB (mid-high) \\
            \bf260701 &  Broad-band noise & - & - & - &- &FB (mid-high) \\
            \bf270702 &  Broad-band noise & - & - & - &- &FB (mid-high) \\
            \bf280703 &   Broad-band noise  & - & - & - &- &FB (mid-high) \\ 
            \hline
            \bf290800 &  FBO + NBO (very blurred) & $8.49$ & $14.58$ & - &- &FB (low) \\
            \bf 300801 &  FBO (blurred) & - & $11.83$ & - &- &FB (low)/NB \\
            \bf310802 &  FBO (blurred) & - & $9.75$ & - & -&FB (low)/NB \\
            \bf320803 &  FBO + NBO  & $4.50$ & $7.41$ & - & -&NB \\

		\hline

	\end{tabular}


	
\end{table*}

\begin{table*}
	\footnotesize
	\centering
	\caption{Timing features found in the PDS extracted for Obs-2. In the column QPO, we refer to any feature which cannot be classified in the three main categories (i.e., NBO, FBO, HBO). These features are generally faint (i.e. barely significant) and difficult to constrain in the model; we report them here for completeness, but they are not shown as part of the main tables of the paper. In this observation, there are no high FB states reported in the HID column. This is because our nomenclature is based on the HID extracted for the full dataset, where the FB is significantly larger than in this observation.}
	\label{app:tab2}
	\begin{tabular}{lccccccc} 
        \hline
	\bf Id  & \bf Timing event & \bf NBO (Hz) & \bf FBO (Hz) & \bf HBO (Hz) &\bf QPO (Hz) & \bf HID \\
       
		\hline
                \bf000000 &  NBO+HBO (hint?) & $6.34$ & - & 21.57&- & NB \\
                
                \bf010001 &  NBO & $6.40$ & - & - &-& NB \\
                
                \bf020002 &  NBO (blurred) & $7.67$ & - & - & - & NB/FB (low) \\
                \hline
                \bf030100 &  FBO (blurred) & - & $13.58$ & - &-& FB (low) \\
                
                \bf040101 &  Broad-band noise & - & - & - &-& FB (low) \\
                
                \bf050102 &  Broad-band noise & - & - & - &-& FB (low-mid) \\
                \hline
                \bf060200 & NBO (blurred) & 4.88 & - & - &-& FB (low) \\
                
                \bf070201 &  Broad-band noise & - & - & - &-& FB (low-mid) \\
                
                \bf080202 &  Broad-band noise & - & - & - &-& FB (mid) \\
                \hline
                \bf090300 &  Broad-band noise & - & - & - &-& FB (low) \\
                
                \bf100301 &  Broad-band noise & - & - & - &-& FB (low) \\
                
                \bf110302 &  Broad-band noise & - & - & - &-& FB (low) \\
                \hline
                \bf120400 &  NBO & $5.84$ & - & - &-& NB \\
                
                \bf130401 &  NBO (very blurred) & $5.81$ & - & - &-& NB \\
                
                \bf140402 &  Broad-band noise & - & - & - &-& NB \\
                \hline
                \bf150500 &  HBO & - & - & $33.66$ &-& HB \\
                
                \bf160501 &  Broad-band noise & - & - & - &-& HB \\
                
                \bf170502 &  Broad-band noise & - & - & - &-& HB \\
                \hline
                \bf180600 &  NBO + HBO (hint?) & $6.80$ & - & $19.50$ &-& NB \\
                
                \bf190601 &  NBO  & $6.68$ & - & - &-& NB \\
                
                \bf200602 &  NBO (blurred) & $7.32$ & - & - &-& NB \\
                \hline
                \bf210700 &  FBO & - & $16.59$ & - &-& FB (low) \\
              
                \bf220701 &  FBO & - & $17.05$ & - &-& FB (low) \\
                
                \bf230702 &  FBO (blurred) + QPO (?) & - & $12.86$ & - &33.44& FB/NB \\
                \hline
                \bf240800 &  NBO & $6.27$ & - & - & -&NB \\
                
                \bf250801 &  NBO & $6.13$ & - & - & -&NB \\
                
                \bf260802 &  NBO & $6.25$ & - & - & -&NB \\
                \hline
                \bf270900 & NBO & $6.46$ & - & - & -&NB \\
                
                \bf280901 &  NBO & $6.43$ & - & - & -&NB \\
                
                \bf290902 &  NBO & $6.20$ & - & - & -&NB \\
                \hline
                \bf301000 & FBO  & - & $16.17$ & - & -&FB (low) \\
                
                \bf311001 &  FBO  & - & $14.22$ & - & -&FB (low) \\
                
                \bf321002 & FBO & - & $13.62$ & - & -&FB (low) \\
                \hline
                \bf331100 & Broad-band noise & - & - & - & -&FB (mid) \\
                
                \bf341101 &  Broad-band noise & - & - & - & -&FB (low-mid) \\
                
                \bf351102 & Broad-band noise & - & - & - & -&FB (mid) \\
                \hline
                \bf361200 & FBO & - &  $15.80$ & - & -&FB (low-mid) \\
                
                \bf371201 & Broad-band noise & - & - & - & -&FB (low-mid) \\
                
                \bf381202 &  Broad-band noise & - & - & - & -&FB (low-mid) \\
                \hline
                \bf391300 & Broad-band noise & - & - & - & -&FB (low-mid) \\
                
                \bf401301 & Broad-band noise & - & - & - & -&FB (low-mid) \\
                
                \bf411302 &  Broad-band noise & - & - & - & -&FB (low-mid) \\
                \hline
                \bf421400 &  Broad-band noise & - & - & - & -&FB (low) \\
                
                \bf431401 &  Broad-band noise & - & - & - & -&FB (low-mid) \\
                
                \bf441402 &  Broad-band noise & - & - & - & -&FB (low-mid) \\

		\hline

	\end{tabular}


	
\end{table*}

\begin{table*}
	\footnotesize
	\centering
	\caption{Timing features found in the PDS extracted for Obs-3. In the column QPO, we refer to any feature which cannot be classified in the three main categories (i.e., NBO, FBO, HBO). These features are generally faint (i.e. barely significant) and difficult to constrain in the model; we report them here for completeness, but they are not shown as part of the main tables of the paper.}
	\label{app:tab3}
	\begin{tabular}{lccccccc} 
        \hline
	\bf Id  & \bf Timing event & \bf NBO (Hz) & \bf FBO (Hz) & \bf HBO (Hz) &\bf QPO (Hz) & \bf HID \\
       
		\hline
                \bf000000 & Broad-band noise & - & - & - &- & FB (low) \\
                
                \bf010001 & Broad-band noise & - & - & - & - &FB (low) \\

                \bf020002 & Broad-band noise & - & - & - & - &FB (low) \\

                \hline

                \bf030100 & NBO  & $6.30$ & - & - & - &NB \\

                \bf040101 & NBO + HBO (hint?) & $6.18$ & - & $23.01$ & - &NB \\

                \bf050102 & NBO & $6.17$ & - & - & - &NB \\

                \hline

                \bf060200 & FBO (blurred)  & - & $9.24$ & - & - &FB(low)/NB \\

                \bf070201 & Broad-band noise & - & - & - & - &FB (low) \\
                
                \bf080201 & Broad-band noise & - & - & - & - &FB (low) \\

                \hline
                
                \bf090300 & Broad-band noise & - & - & - & - &FB (low-mid) \\

                \bf100301 & Broad-band noise  & - & - & - & 66.15 &FB (low) \\

                \bf110302 & FBO (blurred) & - & $16.28$ & - & - &FB (low) \\

                \hline

                \bf120400 & Broad-band noise & - & - & - & - &FB (mid) \\

                \bf130401 & Broad-band noise & - & - & - & - &FB (mid) \\

                \bf140402 & Broad-band noise & - & - & - & - &FB (mid) \\

                \hline

                \bf150500 & Broad-band noise & - & - & - & - &FB (high) \\

                \hline
                
                \bf160600 & FBO  & - & $17.76$ & - & - &FB (low) \\

                \bf170601 & Broad-band noise & - & - & - & - &FB (low) \\

                \bf180602 & Broad-band noise & - & - & - & - &FB (low) \\

                \hline

                \bf190700 & Broad-band noise & - & - & - & - &FB (low) \\

                \bf200701 & Broad-band noise & - & - & - & - &FB (low) \\

                \bf210702 & Broad-band noise & - & - & - & - &FB (low) \\

                \hline

                \bf220800 & Broad-band noise & - & - & - & - &FB (mid) \\

                \bf230801 & FBO & - & $6.83$ & - & - &FB (mid) \\

                \bf240802 & Broad-band noise & - & - & - & - &FB (low-mid) \\

                \hline

                \bf250900 & Broad-band noise & - & - & - & - &FB (low) \\

                \bf260901 & Broad-band noise & - & - & - & - &FB (low) \\

                \bf270902 & Broad-band noise & - & - & - & - &FB (low-mid) \\

                \hline

                \bf281000 & Broad-band noise & - & - & - & - &FB (high) \\

                \bf291001 & Broad-band noise & - & - & - & - &FB (mid-high) \\

                \bf301002 & Broad-band noise & - & - & - & - &FB (mid-high) \\

                \hline

                \bf311100 & Broad-band noise & - & - & - & - &FB (mid-high) \\

                \bf321101 & Broad-band noise & - & - & - & - &FB (mid-high) \\

                \bf331102 & Broad-band noise & - & - & - & - &FB (mid) \\

                \hline

                \bf341200 & FBO  & - & $17.94$ & - & - &FB (low) \\
                
                \bf351201 & FBO  & - & $17.79$ & - & - &FB (low) \\

                \bf361202 & FBO (very blurred) & - & $14.62$ & - & - &FB (low) \\

		\hline

	\end{tabular}


	
\end{table*}

\end{appendix}

\end{document}